\documentclass{llncs}
\usepackage{setspace}
\usepackage{mathdots}

\usepackage{scalerel}

\usepackage{fancybox}

\usepackage[ddmmyy,hhmmss]{datetime}
\usepackage{tikzpagenodes}

\definecolor{myred}{RGB}{165,41,33}
\definecolor{mygrey}{RGB}{130,130,130}

\newcommand{\Nv}{\color{navy}}
\newcommand{\card}[1]{\left|#1\right|}
\definecolor{navy}{rgb}{0,0,0.5}

\newcommand{\dotprecM}{\mathbin{\mbox{$\overset{\raisebox{-1pt}{\Large .}}{\prec}\vphantom{\smash[b]{\prec}}\hspace{-2.7pt}^{M}\,$}}}

\usepackage[english]{babel}

\usepackage{hyperref}
\usepackage{xcolor}
\hypersetup{
    colorlinks = false,
    linkbordercolor = {white},
    citebordercolor = {white},
    urlbordercolor = {white},
}

\usepackage{amsthm}
\usepackage[normalem]{ulem}

\newcommand{\Pow}{{\mathscr{P}}}
\newcommand{\pow}[1]{\Pow({#1})}

\newcommand{\wt}{\,\mathtt{|}\:}

\renewcommand{\leq}{\leqslant}
\renewcommand{\geq}{\geqslant}

\newcommand{\hug}[1]{\left\{\,#1\,\right\}}
\newcommand{\opair}[2]{\left\langle\,#1\,,\:#2\,\right\rangle}
\newcommand{\symm}{\bigtriangleup}

\usepackage[T1]{fontenc}
\usepackage{ae,aecompl}
\usepackage{enumitem}
\usepackage{varwidth}
\usepackage{amsmath}
\usepackage{xspace}
\usepackage{amssymb}
\usepackage{mathtools}
\usepackage{mathrsfs}
\usepackage[ruled]{algorithm}
\usepackage[noend]{algpseudocode}

\newcommand{\COMMENT}[1]{}

\newcommand{\Sec}[1]{Sec.$\:$\ref{#1}}
\newcommand{\Secs}[2]{Sections~$\:$\ref{#1} and \ref{#2}}

\newcommand{\Fig}[1]{Figure~\ref{#1}}

\newcommand{\Vars}[1]{\mathit{Vars}(#1)}

\newcommand{\NP}{\textnormal{\textsf{NP}}\xspace}
\newcommand{\defAs}{\coloneqq}

\newcommand{\dom}[1]{\mathsf{dom}(#1)}
\newcommand{\rk}[1]{\mathsf{rk}\left(#1\right)}

\newcommand{\BSTbb}{\textnormal{$\mathbb{BST}$}\xspace}
\newcommand{\BSTbbPlus}{\textnormal{$\mathbb{BST}^{+}$}\xspace}

\newcommand{\defref}[1]{Def.~\ref{#1}}
\newcommand{\lemsref}[2]{Lemmas~\ref{#1} and \ref{#2}}
\newcommand{\lemref}[1]{Lemma~\ref{#1}}

\newcommand{\corref}[1]{Corollary~\ref{#1}}

\newcommand{\Disj}[1]{\textsc{Disj}(#1)}

\newcommand{\App}[1]{Appendix~\ref{#1}}
\newcommand{\Prop}[1]{Proposition~\ref{#1}}

\def\sqdot{\mathbin{\scalerel*{\strut\rule{1ex}{1ex}}{\cdot}}}

\title{Complexity assessments for decidable fragments of Set Theory.\ IV:\ A quadratic reduction of constraints over nested sets to %
 Boolean formulae %
\thanks{We gratefully acknowledge partial support from project ``STORAGE---Uni\-versit\`{a} degli Studi di Catania, Piano della Ricerca 2020/2022, Linea di intervento 2''.
}}
\author{Domenico Cantone\orcidID{0000-0002-1306-1166}\inst{1} \and \\Andrea De Domenico\orcidID{0000-0002-8973-7011}\inst{2}$\hspace{0.1cm}$\inst{3} \and\\ Pietro Maugeri\orcidID{0000-0002-0662-2885}\inst{1} \and \\Eugenio G. Omodeo\orcidID{0000-0003-3917-1942}\inst{4}}

\institute{Dept.\ of Mathematics and Computer Science, University of Catania, Italy\\
\email{domenico.cantone@unict.it,\ pietro.maugeri@unict.it} \and
Scuola Superiore di Catania, University of Catania, Italy \and
School of Business and Economics, Vrije Universiteit Amsterdam, Netherlands\\
\email{andrea.dedomenico@studium.unict.it} \and
Dept.\ of Mathematics and Earth Sciences, University of Trieste, Italy\\
\email{eomodeo@units.it}}

\begin{document}

\date{}
\maketitle
\thispagestyle{plain}
\begin{abstract}
\noindent As a contribution to quantitative set-theoretic inferencing, a translation is proposed of conjunctions of literals of the
forms $x=y\setminus z$, $x \neq y\setminus z$, and $z =\hug{x}$, where $x,y,z$ stand for variables ranging
over the von Neumann universe of sets, into unquantified Boolean formulae of
a rather simple conjunctive normal form.
The formulae in the target language involve variables ranging
over a Boolean ring of sets, along with a difference operator 
and relators designating equality, non-disjointness and inclusion.
Moreover, the result of each translation is a conjunction
of literals of the forms $x=y\setminus z$, $x\neq y\setminus z$ and of
implications whose antecedents are isolated literals and whose consequents are either
inclusions (strict or non-strict) between variables, or equalities between variables. 

Besides reflecting a simple and natural semantics, which ensures
sat\-is\-fi\-ability-preservation, the proposed translation
has quadratic algorithmic time-complexity, and bridges two languages
both of which are known to have an \NP-complete satisfiability problem.

\medskip

\end{abstract}

\section*{Introduction}

This paper enhances a quadratic-cost method, announced in \cite{journals/fuin/CantoneDMO21} and then presented in \cite{CDMO20}, that translates the formulae of an unquantified language
involving set-theoretic variables, Boolean operators, and membership and equality relators into propositional combinations of purely Boolean literals. The enhancement lies in the availability of a singleton-formation operator `$\hug{\sqdot}$', which increases the expressive power of the source language but whose handling does not worsen the algorithmic time-complexity of the translation; however, the fact that the translation  preserves satisfiability as well as unsatisfiability must be proved anew and differently. The material treated in what follows bridges two complexity taxonomies designed and analyzed in the respective papers \cite{DBLP:journals/tcs/CantoneMO20} and \cite{journals/fuin/CantoneDMO21}, both regarding fragments of set theory that admit satisfiability decision tests.

A fit theoretical framework for the study of our target language is the theory of
Boolean rings, a merely equational first-order theory endowed with finitely many axioms---at times, one blends this theory with an arithmetic of cardinals (see, e.g., \cite{KuncakNR06})---or just a surrogate of it, cf. \Fig{ax:diffAlg}. Frameworks for the study of the source language are such
all-embracing theories as ZF and NBG (the Zermelo-Fraenkel and von~Neumann-Bernays-G\"odel theories), within which one can cast the whole corpus of mathematical disciplines. Boolean algebra is decidable in its entirety (cf. \cite[Sec.~3.7]{RABIN1977595}); ZF is essentially undecidable,
nonetheless an effort to find practical decision algorithms for fragments of it began in 1979. The rationale of this long-standing research is that satifiability testers embodying some knowledge about ZF can act as key inference mechanisms within a programmed system apt to verifying the correctness of large-scale mathematical proofs as envisaged in \cite{SCO11}. When it comes to implementations, complexity emerges as an unescapable issue; this is why we recently undertook (see, 
e.g., \cite{journals/fuin/CantoneDMO21,DBLP:journals/tcs/CantoneMO20}), a systematic study
on the algorithmic complexities of satisfiability testers. 

\emph{A priori}, one would expect the distance between the performances of decision algorithms for
fragments of Boolean algebra, and of the seemingly much more expressive languages whose dictionaries
embody nested membership, to be abysmal. Luckily, though, as we will see, this is not the case. 

\begin{figure}[!htb]
\[\doublebox{$\begin{array}{l|rclcl|ll}
{\bf (D.1)}&\Nv\ x\setminus(y\setminus y) &\Nv=&\Nv x &&&\:&\mbox{Existence of zero} \\\cline{5-7}
{\bf (D.2)}&\Nv\ (x\setminus y)\setminus z&\Nv=&\Nv(x\setminus z)\setminus y&&&&\mbox{Permutativity}\\\cline{5-7}
{\bf (D.3)}&\Nv\ x\setminus(x\setminus y)&\Nv=&\Nv y\setminus (y\setminus x)&&&&\mbox{Commutativity (of intersection)}\\\cline{5-7}
{\bf (D.4)}&\Nv\ (x \setminus y)\setminus y&\Nv=&\Nv x\setminus y&&&& \mbox{Double subtraction}
\end{array}$}\]
\caption{\label{ax:diffAlg}Proper axioms of the variety of \emph{difference algebras}}
\end{figure}

\begin{center}------------\end{center}

We introduce in \Sec{sec:BSTbb} an interpreted formal language, dubbed
\BSTbb, within which one can formulate unquantified Boolean constraints.
Despite its syntax being quite minimal---\BSTbb only encompasses
conjunctions of primitive literals of two forms, namely $x = y \setminus z$ and $x \neq y \setminus z$ ---,
the satisfiability problem for \BSTbb is \NP-complete (see \cite{journals/fuin/CantoneDMO21}). By way of abbreviations,
a number of additional constraints, e.g., literals of the form $x\neq y\cup z$, can be expressed in \BSTbb.

According to our semantics, the domain of discourse to which \BSTbb refers is a universe of nested sets; however,
as will be seen in \Sec{sec:membershipInexpressibility}, every satisfiable propositional combination of \BSTbb literals
(as a special case, a conjunctive \BSTbb constraint) admits a model consisting of sets which are, in a certain sense, ``flat''.
This makes it evident that there is no straight way of expressing membership in \BSTbb.
To detour this drawback, \Sec{sec:expressibility} advances a new notion of expressibility, that embodies an obligation to supply an algorithmic-complexity assessment. Albeit it not being entirely new---in fact, it enhances a similar one proposed in \cite{journals/fuin/CantoneDMO21}---this notion is, we believe, a valuable contribution of this paper and of its companion \cite{CDMO20}.

In terms of the novel notion of expressibility, in \Sec{sec:singletonExpressibility} we will manage to translate a conjunction of literals of the three forms
$x = y \setminus z$, $x \neq y \setminus z$, and $z=\hug{x}$, into a propositional combination of \BSTbb literals.
The proposed translation is, as will be shown in \Secs{sec:satisfiabilityPreservation}{sec:modelXtension}, satisfiability preserving. It leads to a conjunction some of whose
conjuncts are \BSTbb literals, while others are rather simple disjunctions.
The time-complexity of our translation algorithm is quadratic (see \Sec{sec:newTranslationAlg}), which indirectly shows that the satisfiability
problem remains \NP-complete when the singleton operator $\hug{\sqdot}$ is added to the constructs of \BSTbb: this \NP-completeness result
was known (see, e.g., \cite{CantoneOmodeoPolicriti90}), but this paper sheds new light on it. 

\section{The theory \BSTbb}\label{sec:BSTbb}

Boolean Set Theory (\BSTbb) is the quantifier-free theory composed by all conjunctions of literals of the following two types:
\begin{equation}
x = y \setminus z,\qquad x \neq y \setminus z, \label{bstliterals}
\end{equation}
where $x,y$, and $z$ are \emph{set variables} assumed to range over the universe of the well-founded sets.

\smallskip

Semantics for the theory \BSTbb is defined in terms of set assignments. Specifically, given a (finite) collection $V$ of set variables, a \emph{set assignment} $M$ over $V$---the \emph{domain of definition} of $M$, denoted by $\dom{M}$---is any map from $V$ into the \emph{von Neumann universe $\mathcal{V}$} (see below).\footnote{Note that our semantics of \BSTbb does not rely on flat sets of urelements (as would be doable). Working with those would call for minor adjustments, unjustified---and perhaps disturbing---in the economy of this paper.} 
A set assignment $M$ is said to \emph{satisfy}, or to \emph{model}, a given literal $x = y \setminus z$, with $x,y,z \in \dom{M}$, if $Mx = My \setminus Mz$ holds, where `$\setminus$' denotes standard set difference. Likewise, $M$ is said to satisfy the literal $x \neq y \setminus z$ if $Mx \neq My \setminus Mz$ holds. Finally, $M$ satisfies a \BSTbb-conjunction $\varphi$ such that $\Vars{\varphi} \subseteq \dom{M}$ (where $\Vars{\varphi}$ denotes the collection of the variables occurring free in $\varphi$) if it satisfies all of the conjuncts of $\varphi$, in which case we say that $M$ is a \emph{model} of $\varphi$ and write $M \models \varphi$. A \BSTbb-conjunction is said to be \emph{satisfiable} if it has some model, otherwise it is said to be \emph{unsatisfiable}. 

In \cite{journals/fuin/CantoneDMO21}, it is proved that the satisfiability problem for \BSTbb, namely the problem of establishing algorithmically the satisfiability status of any given \BSTbb-conjunction, is \NP-complete.

We shall also be interested in the extension \BSTbbPlus of the theory \BSTbb consisting of all propositional combinations (resulting from unrestrained use of the logical connectives $\wedge$,$\vee$,$\longrightarrow$,$\longleftrightarrow$,$\neg$) of atomic formulae of type $x = y \setminus z$. It is not hard to check that the satisfiability problem for \BSTbbPlus can be reduced to the satisfiability problem for \BSTbb in nondeterministic polynomial time, and therefore it is \NP-complete in its turn.

\subsection{The von Neumann universe}

We recall that the von Neumann universe $\mathcal{V}$ of (well-founded) sets, also dubbed \emph{von Neumann cumulative hierarchy}, is built up through a transfinite sequence of steps as the union $\mathcal{V} \defAs \bigcup_{\alpha \in \mathit{On}} \mathcal{V}_{\alpha}$ of the levels $\mathcal{V}_{\alpha} \defAs \bigcup_{\beta < \alpha} \pow{\mathcal{V}_{\beta}}$, with $\pow{\sqdot}$ denoting the powerset operator and $\alpha$ ranging over the class $\mathit{On}$ of all ordinals.

Based on the level of first appearance in the von Neumann hierarchy, one can define the rank of any set $s$, denoted $\rk s$. Specifically, $\rk s$ is the ordinal $\alpha$ such that $s \in \mathcal{V}_{\alpha + 1} \setminus \mathcal{V}_{\alpha}$. Hence, for every $\alpha \in \mathit{On}$, the set $\mathcal{V}_{\alpha + 1} \setminus \mathcal{V}_{\alpha}$, hereinafter denoted $\mathcal{V}^{\#}_{\alpha}$, collects all sets having rank $\alpha$.

The following lower bound on the number of well-founded sets of any positive integer rank $n$,
to be proved as \Prop{prop:sizeVsharp} in \App{app:wasLemmaOne}, will be useful:

\[
\card{\mathcal{V}^{\#}_{n}}\geq 2^{n-1}.
\]

Some handy properties of the rank function that we shall tacitly use are the following, which hold for all sets $s,t \in \mathcal{V}\,$:
\begin{enumerate}[label=$\bullet$]
\item if $s \in t$ then $\rk s < \rk t$;

\item if $s \subseteq t$ then $\rk s \leq \rk t$;

\item $\rk s = \begin{cases}
0 & \text{if } s = \emptyset\,,\\
\sup_{u \in s} \big(\rk u + 1\big) & \text{otherwise.}
\end{cases}
$
\end{enumerate}

\smallskip

We also recall that well-foundedness, as enforced by the \emph{regularity} or \emph{foundation axiom} of Zermelo-Fraenkel set theory, precludes the formation of infinite descending membership chains of the form
\[
\cdots \in s_{2} \in s_{1} \in s_{0},
\]
and in particular of membership cycles
\[
s_{0} \in s_{n} \in \cdots \in s_{2} \in s_{1} \in s_{0},
\]
for any sequence $s_{0},s_{1},s_{2},\ldots$ of sets.

\subsection{Existential expressibility and $\mathcal{O}(f)$-expressibility}\label{sec:expressibility}

In spite of the parsimony of \BSTbb as just presented, it turns out (see \cite{journals/fuin/CantoneDMO21}) that several other Boolean constructs, such as the ones in the list of literals
\begin{equation}
\begin{aligned}\label{bstexreliterals}
\phantom{\!\!\!\!\!\!}&x = \varnothing,\qquad x \subseteq y,\qquad x = y \cap z,\qquad x = y \cup z,\qquad \!\!\!\phantom{\neg}\Disj{x,y}, \qquad \!\!\!x \subsetneq y,\\
\phantom{\!\!\!\!\!\!}&x \neq \varnothing,\qquad x \nsubseteq y,\qquad x \neq y \cap z,\qquad x \neq y \cup z,\qquad \!\!\!\neg\Disj{x,y},
\end{aligned}
\end{equation}
can be expressed existentially in \BSTbb, where $\Disj{a,b}$ is a short for $a \cap b = \varnothing$. 

\bigskip

Formally, \emph{existential expressibility} is defined as follows (cf.\ \cite{journals/fuin/CantoneDMO21}, wherein several applications of this notion are reported).

\begin{definition}[Existential expressibility]\label{express}
A formula $\psi(\vec x)$ is said to be \emph{existentially expressible} in a theory $\mathcal{T}$ if there exists a $\mathcal{T}$-formula $\Psi(\,\vec x, \vec z\,)$ such that 
\[
\models\ \ \psi(\,\vec x\,) \: \longleftrightarrow \: (\,\exists \vec z\,)\; \Psi(\,\vec x, \vec z\,),
\]
where $\vec x$ and $\vec z$ stand for tuples of set variables.
\end{definition}

Existential expressibility has been generalized in \cite{journals/fuin/CantoneDMO21} into the definition of $\mathcal{O}(f)$-expressibility. The latter notion enabled, in \cite{journals/fuin/CantoneDMO21}, a detailed complexity taxonomy of the subfragments of \BSTbb.

Here we slightly generalize $\mathcal{O}(f)$-expressibility so that it copes with collections $\mathcal{C}$ of formulae, rather than with single formulae as its original definition did; another difference lies in the fact that the generalized notion has to do
with a source theory $\mathcal{T}_1$ and a target theory $\mathcal{T}_2$, whereas \cite{journals/fuin/CantoneDMO21} took it for granted that
source and target were the same.

\begin{definition}[$\mathcal{O}(f)$-expressibility]\label{weakExprss}
Let $\mathcal{T}_{1}$ and $\mathcal{T}_{2}$ be any theories and $f\colon \mathbb{N} \rightarrow \mathbb{N}\,$ be a given map. A collection $\mathcal{C}$ of formulae is said to be \emph{$\mathcal{O}(f)$-expressible from $\mathcal{T}_{1}$ into $\mathcal{T}_{2}$} if there exists a map
\begin{equation}
\opair{\varphi(\,\vec y\,)}{\psi(\,\vec x\,)}\ \mapsto\ \Xi^{\psi}_{\varphi}(\,\vec x, \vec y, \vec z\,) \label{weakExpressmap}
\end{equation}
from $\mathcal{T}_{1} \times \mathcal{C}$ into $\mathcal{T}_{2}$, where no variable in $\vec z$ occurs in either $\vec x$ or $\vec y$, such that the following conditions are satisfied:
\begin{enumerate}[label=(\alph*)]
\item\label{weakExprssN} the mapping~(\ref{weakExpressmap}) can be computed in $\mathcal{O}\big(f(\card{\varphi \wedge \psi})\big)$-time,

\item\label{weakExprssA} if $\varphi(\,\vec y\,) \wedge \Xi^{\psi}_{\varphi}(\,\vec x, \vec y, \vec z\,)$ is satisfiable, so is $\varphi(\,\vec y\,) \wedge \psi(\,\vec x\,)$,

\item\label{weakExprssB} $\models \big(\varphi(\,\vec y\,) \:\wedge\: \psi(\,\vec x\,)  \big) \:\longrightarrow\: (\,\exists \vec z\,)\Xi^{\psi}_{\varphi}(\,\vec x, \vec y, \vec z\,)$.
\end{enumerate}
\end{definition}

The main results in this paper are that atoms of the form $z =\hug{ x }$ are not existentially expressible in \BSTbbPlus, whereas any conjunction of such atoms is $\mathcal{O}(n^{2})$-expressible from \BSTbb into \BSTbbPlus.

\section{Existential inexpressibility of $z =\hug{ x }$ in \BSTbbPlus}\label{sec:membershipInexpressibility}

If atoms of the form $z =\hug{ x }$ were existentially expressible in \BSTbbPlus,
membership atoms $x\in y$ would also be expressible; in fact, the presence of `$\hug{\sqdot}$' permits downgrading `$\in$' from primitive to derived construct:
\begin{lemma}\label{lem:memb_singl} $\models x \in y \longleftrightarrow  (\exists z)( z = \hug{x} \wedge z \subseteq y)\,,$ if $x,y,z$ are distinct variables.
\end{lemma}
\begin{proof}
If $M \models z = \hug{x} \wedge z \subseteq y$ holds, then $M \models x \in y$, since $Mx \in Mz \subseteq My\,$.

Conversely, if $M\models x \in y$ holds, extend $M$ by putting $Mz = \hug{Mx}$; then $Mx \in My$
yields $M \models z = \hug{x} \wedge  z \subseteq y$, so that $M \models (\exists z)(z = \hug{x} \wedge  z \subseteq y)\,$.
\end{proof}
In this section we show that membership atoms $x\in y$ are not existentially expressible in \BSTbbPlus; therefore, by the lemma just seen, atoms $z =\hug{ x }$ are not existentially expressible in \BSTbbPlus either. Specifically, we will prove that every satisfiable \BSTbbPlus-formula $\Psi$ admits a ``flat'' model $M$, namely a model whose union $\bigcup_{v \in \Vars{\Psi}} Mv$ of values is made up of members all of the same positive rank. Consequently, $Mx \notin My$ for any $x,y \in \Vars{\Psi}$, and hence $M \not\models x \in y\,$.

\begin{definition}
For every ordinal $\,\flat \geq 1$, a set assignment $M$ over a collection $V$ of variables is said to be \emph{$\flat$-flat} if all sets in 
the \emph{realm} $\bigcup \hug{\,Mv \wt v \in V\,}$ of $M$ have rank $\flat$.
\end{definition}

No membership atom $x \in y$ is satisfied by any $\flat$-flat set assignment:

\begin{lemma}\label{notin}
Let $M$ be a $\,\flat$-flat set assignment over a collection $V$ of variables. Then $Mx \notin My$, for any $x,y \in V$.
\end{lemma}
\begin{proof}
Because of the $\flat$-flatness of $M$, for every $x \in V$ either $\rk {Mx} = 0$ (when $Mx = \emptyset$) or $\rk {Mx} = \flat + 1$ (when $Mx \neq \emptyset$). Hence, in any case $\rk {Mx} \neq \flat$ (since $\flat$-flatness presupposes $\flat \geq 1$), and therefore $Mx \notin \bigcup \hug{My \mid y \in V }$.
\end{proof}

A satisfiable \BSTbbPlus-formula always admits a $\flat$-flat model, for sufficiently large $\flat$. This is proved in the next lemma.

\begin{lemma}\label{lemma:rho-flat}
Every satisfiable formula $\Phi$ of \BSTbbPlus admits a $\flat$-flat set model, for any $\flat \geq \card{\Vars{\Phi}} + 1$.
\end{lemma}
\begin{proof}
Let $\Phi$ be a satisfiable formula of \BSTbbPlus, and let $M$ be a set model of $\Phi$. Let $\phi_{M}^{+}$ be the conjunction of all distinct atoms $x = y \setminus z$ occurring in $\Phi$ that are satisfied by $M$. Likewise, let $\phi_{M}^{-}$ be the conjunction of all distinct literals $x \neq y \setminus z$ satisfied by $M$ and such that $x = y \setminus z$ occurs in $\Phi$. Finally, let 
\begin{equation}\label{def_varphi}
\phi_{M} \defAs \phi_{M}^{+} \wedge \phi_{M}^{-}.
\end{equation}
Plainly, $M$ satisfies $\phi_{M}$ by construction. Additionally, by propositional reasoning, every set model of $\phi_{M}$ satisfies our initial formula $\Phi$. Thus, it is enough to show that the conjunction $\phi_{M}$ admits a $\flat$-flat set model for every $\flat \geq n+1$, where $n \defAs \card{\Vars{\phi_{M}}} = \card{\Vars{\Phi}}$.

We prove that $\phi_{M}$ admits a $\flat$-flat set model by contracting each nonempty region $R_{W}$ of $M$ of the form
\[
R_{W} \defAs \left(\bigcap \hug{Mx \mid x \in W }\right) \setminus \bigcup \hug{My \mid y \in \Vars{\phi_{M}} \setminus W },
\]
for $\emptyset \neq W \subseteq \Vars{\phi_{M}}$, into a distinct singleton of rank $\flat + 1$ (hence containing a single member of rank $\flat$).

Since the map $\kappa \mapsto 2^{\kappa} - \kappa$ is strictly increasing for $\kappa \geq 1$, for every integer $\flat \geq n+1$ we have
\[
\card{\mathcal{V}^{\#}_{\flat}} = \card{\mathcal{V}_{\flat+1}} - \card{\mathcal{V}_{\flat}} = 2^{\card{\mathcal{V}_{\flat}}} - \card{\mathcal{V}_{\flat}} \geq 2^{\card{\mathcal{V}_{n+1}}} - \card{\mathcal{V}_{n+1}} = \card{\mathcal{V}_{n+2}} - \card{\mathcal{V}_{n+1}} = \card{\mathcal{V}^{\#}_{n+1}} \geq 2^{n},
\]
where the latter inequality follows from \Prop{prop:sizeVsharp} (see \App{app:wasLemmaOne}).
Hence, there exists an injective map $\Im_{\flat} \colon \pow{\Vars{\phi_{M}}} \rightarrow \mathcal{V}^{\#}_{\flat}$ from the collection of the nonempty subsets of $\Vars{\phi_{M}}$ into the family $\mathcal{V}^{\#}_{\flat}$ of the (hereditarily finite) sets of rank $\flat$.

Next, we define a set assignment $M^{*}_{\flat}$ over $\Vars{\phi_{M}}$ by putting
\[
M^{*}_{\flat} x \defAs \hug{ \Im_{\flat}(W) \wt  x \in W \subseteq \Vars{\phi_{M}} \wedge R_{W} \neq \emptyset}.
\]
By construction, the assignment $M^{*}_{\flat}$ is $\flat$-flat. In addition, it is not hard to check that, for every $\emptyset \neq W \subseteq \Vars{\phi_{M}}$, the region $R^{*}_{W}$ of $M^{*}$ defined by
\[
R^{*}_{W} \defAs \left(\bigcap \hug{M^{*}_{\flat}x \wt  x \in W }\right) \setminus \bigcup \hug{M^{*}_{\flat}y \wt  y \in \Vars{\phi_{M}} \setminus W }
\]
is nonempty if and only if so is its corresponding region $R_{W}$ of $M$. Thus, $M^{*}_{\flat}$ satisfies $\phi_{M}$.
\end{proof}

We are now ready to prove that membership atoms $x \in y$---and hence, by \lemref{lem:memb_singl}, atoms of the form $z=\hug{x}$---are not existentially expressible in \BSTbbPlus.

\begin{theorem}
The atom $x \in y$ is not existentially expressible in \BSTbbPlus.
\end{theorem}
\begin{proof}
By way of contradiction, assume that $x \in y$ is existentially expressible by a formula $\Psi(x,y,\vec{z})$ of \BSTbbPlus involving only atoms of type $x' = y' \setminus z'$. Hence,
\begin{equation}\label{expressibility}
\models x \in y \longleftrightarrow (\exists \vec{z})\;\Psi(x,y,\vec{z})
\end{equation}
must hold.

Since $x \in y$ is trivially satisfiable, so are---by \eqref{expressibility}---$(\exists \vec{z})\;\Psi(x,y,\vec{z})$ and $\Psi(x,y,\vec{z})$. Thus, by \lemref{lemma:rho-flat}, $\Psi(x,y,\vec{z})$ is modeled by a $\flat$-flat set assignment $M^{*}$; hence, by \lemref{notin}, $M^{*} x \notin M^{*} y$ holds. It follows that $M^{*} \models x \notin y\wedge (\exists \vec{z})\;\Psi(x,y,\vec{z}) $ holds, and therefore $M^{*} \not\models (\exists \vec{z})\;\Psi(x,y,\vec{z}) \longrightarrow x \in y$, which contradicts \eqref{expressibility}.
\end{proof}

\section{$\mathcal{O}(n^2)$-expressibility in \BSTbbPlus of singleton-atom conjunctions}\label{sec:singletonExpressibility}

Conforming with \defref{weakExprss}, we shall prove that any conjunction $\psi(\,\vec x\,)$ 
of atoms of the form $x=\hug{y}$ is $\mathcal{O}(n^{2})$-expressible from \BSTbb into \BSTbbPlus by exhibiting a map 
\[
\opair{\varphi(\,\vec y\,)}{\psi(\,\vec x\,)}\ \mapsto\ \Xi^{\psi}_{\varphi}(\,\vec x, \vec y, \vec z\,),
\]
which can be computed in quadratic time and such that conditions \ref{weakExprssA} and \ref{weakExprssB} of \defref{weakExprss} are satisfied, where $\varphi(\,\vec y\,)$ ranges over the collection of  \BSTbb-conjunctions and the variables in $\vec z$ are distinct from those in $\vec x$ and in $ \vec y$.

Thus let $\varphi(\vec y)$ be any \BSTbb-conjunction and $\psi(\vec x)$ be of the said form.
For each variable $x \in \Vars{\varphi \wedge \psi}$ we introduce a new distinct auxiliary variable $\tilde{x}$ (these variables will enforce that $x\in y$ only if $\tilde{x}\subsetneq\tilde{y}$). Then we put:
\begin{align*} 
\Xi_\varphi^\psi \coloneqq &\ \;\bigwedge_{x = \hug{y} \text{ in } \psi}
 x \nsubseteq y%
 \ \wedge\\[0.25cm]
&\bigwedge_{\substack{x = \hug{y} \text{ in } \psi\\ x' = \{y'\} \text{ in } \psi}}\big( y = y' \longleftrightarrow x = x'\big)\ \wedge
\bigwedge_{\substack{x = \hug{y} \text{ in } \psi\\v \in \Vars{\varphi\wedge\psi}}}\big(\neg\Disj{x,v}\longrightarrow x \subseteq v \big)\ \wedge\\[0.3cm]
&\bigwedge_{\substack{x = \hug{y} \text{ in } \psi\\v \in \Vars{\varphi\wedge\psi}}}\big(\neg\Disj{x,v} \longrightarrow \tilde{y} \subsetneq \tilde{v}\big)\ \wedge
\bigwedge_{x,y \in \Vars{\varphi\wedge\psi}}\big( x = y \longrightarrow \tilde{x} = \tilde{y}\big)
\end{align*}
(thus, the list $\vec z$ of variables in \defref{weakExprss} is the collection $\widetilde{\vec x}$ of all auxiliary set variables $\tilde{x}$).

Plainly, $\Xi^{\psi}_{\varphi}$ is a \BSTbbPlus-formula---in fact, it is a conjunction of a rather simple form---, which
satisfies the following proposition, implying condition \ref{weakExprssN} of \defref{weakExprss}:

\begin{lemma}\label{quadraticCom}
$\Xi^{\psi}_{\varphi} = \Theta\left( \card{\Vars{\varphi \wedge \psi}}^{2}\right)$. \qed
\end{lemma}

The proof of this lemma is postponed to \Sec{sec:newTranslationAlg}.
In the next two subsections, we shall prove that 
\begin{enumerate}[label=$\bullet$]
\item if $\varphi(\,\vec y\,) \wedge \Xi^{\psi}_{\varphi}(\,\vec x, \vec y, \widetilde{\vec x}\,)$ is satisfiable, then so is $\varphi(\,\vec y\,) \wedge \psi(\,\vec x\,)$, and

\item every model of $\varphi(\,\vec y\,) \wedge \psi(\,\vec x\,)$ can be extended into a model of $\Xi^{\psi}_{\varphi}(\,\vec x, \vec y, \widetilde{\vec x}\,)$,
\end{enumerate}
thus showing that also conditions \ref{weakExprssA} and \ref{weakExprssB} of \defref{weakExprss} are fulfilled, which proves that every singleton-atom conjunction is $\mathcal{O}(n^{2})$-expressible from \BSTbb into \BSTbbPlus.

\subsection*{Translation examples}

Here we digress to provide a few examples illustrating how the conjunction $\Xi_\varphi^\psi$ renders the formula $\varphi \wedge \psi$.

Singletons are characterized by the fact they contain exactly one element, that is why both
\[\bigwedge_{x = \hug{y} \text{ in } \psi} x\nsubseteq y \ \ \ %
\text{and}\ \ %
\bigwedge_{\substack{x = \hug{y} \text{ in } \psi\\v \in \Vars{\varphi\wedge\psi}}}\big(\neg\Disj{x,v}\longrightarrow x \subseteq v \big)
\]
(the former of which entails $x \neq \varnothing$) occur in our translation.
Moreover the singleton-atom $x =\hug{y}$ plainly implies $y \in x$, therefore the ordering relation induced by $\in$ must be preserved.\footnote{See, below, the beginning of \Sec{sec:modelXtension}.} This is the rationale for
including, in our translation, the constraints
\[
\bigwedge_{\substack{x = \hug{y} \text{ in } \psi\\v \in \Vars{\varphi\wedge\psi}}}\big(\neg\Disj{x,v} \longrightarrow \tilde{y} \subsetneq \tilde{v}\big)\ \ \ %
\text{and}\ \ %
\bigwedge_{x,y \in \Vars{\varphi\wedge\psi}}\big( x = y \longrightarrow \tilde{x} = \tilde{y}\big)\,.
\]

\begin{example}
The conjunction $\varphi \wedge \psi$, where \[\varphi \coloneqq a = b \setminus c\qquad\mbox{and}\qquad 
\psi \coloneqq x = \hug{y}\ \wedge\ y = \{z\}\ \wedge\ z = \{x\},
\]
is plainly unsatisfiable because it implies the membership cycle $x \in z \in y \in x$.
To reflect this, $\Xi_\varphi^\psi$ encompasses the literals and implications
\begin{align*}
x\neq\varnothing\,,\ \ \big(\neg\Disj{x,x} \longrightarrow \tilde{y} \subsetneq \tilde{x}\big)\,,\\
y\neq\varnothing\,,\ \ \big(\neg\Disj{y,y} \longrightarrow \tilde{z} \subsetneq \tilde{y}\big)\,,\\
z\neq\varnothing\,,\ \ \big(\neg\Disj{z,z} \longrightarrow \tilde{x} \subsetneq \tilde{z}\big)\,,
\end{align*}
showing that it is unsatisfiable due to the cycle $\tilde{x} \subsetneq \tilde{z} \subsetneq \tilde{y} \subsetneq \tilde{x}$.
\end{example}

\begin{example}
The  conjunction $\varphi \wedge \psi\,$, where $\varphi \coloneqq y = x \setminus z$ and $\psi \coloneqq x = \hug{y} \wedge y = \{z\}$, is not satisfiable; in fact any set assignment $M$ satisfying this formula is such that $My = Mx \setminus Mz\,$, $Mx = \{My\}$, and $My = \{Mz\}$; thus $My = Mz$, and hence $My = \hug{My}$, a contradiction.

Our translation $\Xi_\varphi^\psi$ comprises:
\[
y \neq \varnothing\ \wedge\ x \nsubseteq y\ \wedge\ \big(\neg\Disj{x,y} \longrightarrow x \subseteq y\big).
\]
Any model $M$ for $\varphi \wedge \Xi_\varphi^\psi$ is such that, $My \subseteq Mx$ and $My \neq \emptyset$ hold, so that $\neg\Disj{Mx,My}$; but then $Mx \subseteq My$ must hold, which conflicts with $x \nsubseteq y\,$.
\end{example}

As we will prove in \Sec{sec:modelXtension}, satisfiability carries over from $\varphi \wedge \psi$ to $\varphi \wedge \Xi_\varphi^\psi$. Here is an interesting example of this:

\begin{example}
The conjunction $\varphi \wedge \psi$, where
\[
\varphi \coloneqq x = y \setminus y'\ \wedge\ x = z \setminus z',\quad \text{and}\quad \psi \coloneqq z = \{y\},
\]
is satisfiable. In fact, $\varphi \wedge \psi$ is plainly satisfied by every set assignment $M$ over $\Vars{\varphi \wedge \psi}$ of the form
\begin{equation}\label{form}
Mx \defAs \emptyset, \quad My \defAs s, \quad Mz \defAs \hug{s}, \quad 
My' \defAs s \cup s', \quad Mz' \defAs \hug{s} \cup s'',
\end{equation}
where $s$, $s'$, and $s''$ are any well-founded sets.

It can easily be checked that, for every set assignment $M$ of the form \eqref{form}, the extension $M^{+}$ of $M$ over the auxiliary variables $\tilde{v}$, for $v \in \Vars{\varphi \wedge \psi}$ and where
\[
M^{+} \tilde{v} \defAs \begin{cases}
s \cup \hug{s} & \text{if } s \in Mv\\
\emptyset & \text{otherwise}, 
\end{cases}
\]
satisfies $\Xi_\varphi^\psi$, so that $M \models (\exists \,\tilde{x},\tilde{y},\tilde{z},\tilde{y}',\tilde{z}' ) \Xi_\varphi^\psi$ holds. Thus, to show that condition \ref{weakExprssB} of \defref{weakExprss} is satisfied, namely that $\models \big(\varphi \:\wedge\: \psi  \big) \:\longrightarrow\: 
(\exists \,\tilde{x},\tilde{y},\tilde{z},\tilde{y}',\tilde{z}' ) \Xi_\varphi^\psi$ holds, it is enough to check that the conjunction $\varphi \wedge \psi$ is satisfied by set assignments of the form \eqref{form} only. Let then $\overline M$ be any model for $\varphi \wedge \psi$. Since $\overline M x \subseteq \overline M z$ and $\overline M z = \hug{\overline M y}$, either $\overline M x = \emptyset$ or $\overline M x = \hug{\overline M y}$ holds. The latter case can be readily ruled out, for in view of $\overline M x \subseteq \overline M y$ it would follow $\overline M y \in \overline M y$, which is untenable in the realm of well-founded sets. Thus, $\overline M x = \emptyset$ must hold. Letting $s \defAs \overline M y$---so that $\overline M z = \hug{s}$---, since $\overline M y \subseteq \overline M y'$ and $\overline M z \subseteq \overline M z'$, we have $\overline M y' = s \cup s'$ and $\overline M z' = \hug{s} \cup s''$, for some sets $s'$ and $s''$, and therefore $\overline M$ has the form \eqref{form}.
\end{example}

\subsection{If $\varphi \wedge \Xi_\varphi^\psi$ is satisfiable, then so is $\varphi \wedge \psi$}\label{sec:satisfiabilityPreservation}

In order to convert a model $M^{(0)}$ of $\varphi \wedge \Xi_\varphi^\psi$ into a
model of $\varphi \wedge \psi$, we can assume that $M^{(0)}$ is $\flat$-flat 
over $\Vars{\varphi \wedge \Xi_\varphi^\psi}$ for some integer $\flat>\card{\Vars{\varphi\wedge\Xi_\varphi^\psi}}$: on the one hand,
\lemref{lemma:rho-flat} enables us to do so; on the other hand, \lemref{notin} tells
us that such an $M^{(0)}$ does not model any of the atoms $x=\hug{y}$ in $\psi$.
For $i=1,\dots,m$ (where $m$ is a suitable integer, $0\leq m\leq\card{\hug{\mbox{atoms in $\psi$}}}$), we will
lift $M^{(i-1)}$ into a model $M^{(i)}$ of $\varphi \wedge \Xi_\varphi^\psi$
such that the atoms in $\psi$ modeled by $M^{(i-1)}$ remain true in $M^{(i)}$, while
at least one conjunct $x_i=\hug{y_i}$ of $\psi$, false in $M^{(i-1)}$,
becomes true in $M^{(i)}$. This iterative process will end as soon as all atoms in $\psi$ are
true; actually, if $\psi_i$ is the conjunction of  those atoms in $\psi$ which $M^{(i)}$
makes true for the first time, 
$\psi_1 \wedge \cdots \wedge \psi_m$ will coincide with $\psi\,$---up to an atom rearrangement.

In our set up, each $\psi_i$ must hence comprise at least a conjunct $x_i=\hug{y_i}$ of $\psi$ not
appearing in any $\psi_j$ with $j<i$. The selection of $x_i=\hug{y_i}$
will shape the transformation of $M^{(i-1)}$ into $M^{(i)}$ and will be based on the criterion that  
the atom in question be minimal, in $\psi$ (deprived of all atoms in $\psi_1 \wedge \cdots \wedge \psi_{i-1}$),
in regard to the ordering $\prec^{M^{(0)}}$ induced on the conjuncts of $\psi$ as specified below:
\begin{definition}\label{rel}
Given a model $M$ of $\varphi \wedge \Xi_\varphi^\psi$, we put $\ell \dotprecM \ell'$ for all atoms $\ell \coloneqq x \!\!=\!\! \hug{\hspace{-2.5pt}y\hspace{-2.5pt}}$ and $\ell' \coloneqq x' \!\!=\!\! \hug{\hspace{-2.5pt}y'\hspace{-2.5pt}}$ in $\psi$ such that $\neg\Disj{Mx,My'}$ holds. Then, for all conjuncts $\ell$ and $\ell'$ in $\psi$, we put $\ell \prec^{M} \ell'$ provided that
\[
\ell = \ell_{0} \dotprecM \ell_{1} \dotprecM \cdots \dotprecM \ell_{n} = \ell',
\]
for some $\ell_{0}, \ell_{1}, \ldots , \ell_{n}$ in $\psi$ with $n \geq 1$. 
\end{definition}

\noindent This definition enforces that

\begin{lemma}\label{rel:parorder}
The relation $\prec^M$ is a strict partial order.
\end{lemma}
\begin{proof}
Plainly, $\prec^M$ is a transitive relation; it will hence suffice to prove that there is no cycle $\ell_0 \prec^M \cdots \prec^M \ell_{n} = \ell_0$.
Assuming, by way of contradiction, that such a cycle exists, there would be
atoms $\ell'_{0},\ell'_{1},\ldots,\ell'_{m}$ in $\psi$ such that
\[
\ell'_{0} \dotprecM \ell'_{1} \dotprecM \cdots \dotprecM \ell'_{m-1} \dotprecM \ell'_{m} = \ell'_{0},
\]
where $\ell'_{i}$ has the form $x_{i} \!\!=\!\! \hug{\hspace{-2.5pt}y_{i}\hspace{-2.5pt}}$ for $i=0,1,\ldots,m$. Then, we would have \[\neg \Disj{Mx_{0},My_{1}}, \ldots, \neg \Disj{Mx_{m-1},My_{0}}\,.\] Since $M \models \Xi_\varphi^\psi$, then in particular 
\[
M \models \bigwedge_{\substack{x = \hug{y} \text{ in } \psi\\v \in \Vars{\varphi\wedge\psi}}}\big(\neg\Disj{x,v} \longrightarrow \tilde{y} \subsetneq \tilde{v}\big)\, ,
\]
and therefore
\[
M\tilde{y}_{0} \subsetneq M\tilde{y}_{1} \subsetneq \cdots \subsetneq M\tilde{y}_{m-1} \subsetneq M\tilde{y}_{0}\, ,
\]
thus yielding the contradiction $M\tilde{y}_{0} \subsetneq M\tilde{y}_{0}$.

The acyclicity of $\prec^{M}$ entails its irreflexivity, whence the claim follows.
\end{proof}

Let $M$ be a set assignment over the variables of $\varphi \wedge\Xi_\varphi^\psi$, and $x = \hug{y}$ be any atom in $\psi$ such that 
\begin{equation}\label{notincond}
\forall v \in \Vars{\varphi \wedge \psi}\:\big(\,My \notin Mv\,\big)\,.
\end{equation}

\noindent We define the set assignment $M_{x,y}$ by putting, for all $v\in\Vars{\varphi\wedge\psi}$:

\begin{equation}\label{functionM}
M_{x,y}v \coloneqq  \begin{cases}
Mv &\text{if } \Disj{Mx,Mv}\,,\\
\big(\,Mv\setminus Mx\,\big)\uplus\hug{My} &\text{otherwise,}
\end{cases}
\end{equation}
\noindent and
\begin{equation}\label{functionMpp}M_{x,y}\tilde{v} \coloneqq M\tilde{v}\,.
\end{equation} 
\noindent
(As usual, $\uplus$ denotes disjoint set union.) 

It turns out that the set assignments $M_{x,y}$ and $M$ model the very same literals of type $u \subseteq v$ and $\Disj{u,v}$, with $u,v \in \Vars{\varphi \wedge \psi}$. This is proved in the next two lemmas.
\begin{lemma}\label{equallemma}
Let $x = \hug{y}$ be an atom in $\psi$ and $M$ be a model of $\varphi \wedge \Xi_\varphi^\psi$ satisfying condition \eqref{notincond}; then $Mu \subseteq Mv \longleftrightarrow M_{x,y}u \subseteq M_{x,y}v\,$,
for all $u,v \in \Vars{\varphi\wedge\psi}$.
\end{lemma}
\begin{proof}
Plainly $Mu \subseteq Mv \longrightarrow M_{x,y}u \subseteq M_{x,y}v$. To get the converse, suppose that $M_{x,y}u \subseteq M_{x,y}v$ and $s \in Mu\,$:
we must prove that $s\in Mv$. On the one hand,
if $s \notin Mx$ then $s \in Mu\setminus Mx\subseteq M_{x,y}u\subseteq M_{x,y}v$
and $s\neq My$ hold, entailing $s\in Mv$. On the other hand,  if $s \in Mx$ holds, then $\neg\Disj{Mx,Mu}$ follows from $s\in Mx\cap Mu$, and therefore $My\in M_{x,y}u\subseteq M_{x,y}v$ and $\neg\Disj{Mx,Mv}$ hold. Since $M\models \Xi_\varphi^\psi$, we then have $Mx\subseteq Mv$, and thus $s\in Mv$ again, as sought. %
\end{proof}

\begin{lemma}\label{disjlemma}
Let $x = \hug{y}$ be an atom in $\psi$ and $M$ be a model of $\varphi \wedge \Xi_\varphi^\psi$ satisfying \eqref{notincond}; then $\Disj{Mu,Mv} \!\longleftrightarrow\! \Disj{M_{x,y}u,M_{x,y}v}$ holds
for all $u,v \in \Vars{\varphi\wedge\psi}$.
\end{lemma}

\begin{proof}
Plainly $\neg\Disj{Mu,Mv} \longrightarrow \neg\Disj{M_{x,y}u,M_{x,y}v}$. To get the converse, suppose that $s\in M_{x,y}u \cap M_{x,y}v\:$: we must prove that $Mu\cap Mv\neq\emptyset\,$. If $s\neq My$, then $s\in Mu\setminus Mx$ and $s\in Mv\setminus Mx$; so we are done. Otherwise, $Mx$ intersects both of $Mu$ and $Mv$; but then,
since $M\models \Xi_\varphi^\psi$, we get $\emptyset \neq Mx \subseteq Mu \cap Mv$.\end{proof}

\begin{lemma}\label{oneexpress}
Let $x = \{y\}$ be an atom in $\psi$ and $M$ be a model of $\varphi \wedge \Xi_\varphi^\psi$ satisfying \eqref{notincond}.
Then $M_{x,y}\models\varphi \wedge \Xi_\varphi^\psi \wedge x = \{y\}$ and $M_{x,y}y=My$, so that $M_{x,y}x\neq M x$.
\end{lemma}
\begin{proof}

To prove that $M_{x,y} \models \varphi$, we will get $M_{x,y}\models\phi$
from $M\models\phi$ for each conjunct $\phi$ of $\varphi$.

Consider first a conjunct $u = v \setminus w$ of $\varphi$.
By \lemsref{equallemma}{disjlemma}, $M_{x,y}u \subseteq M_{x,y}v \setminus M_{x,y}w$ follows from $Mu = Mv \setminus Mw$. Supposing then that $s\in M_{x,y}v \setminus M_{x,y}w$, we must prove that $s\in M_{x,y}u$. In fact, either $s\in 
Mv\setminus Mx$ or $s=My$ and $Mx\cap Mv\neq\emptyset$ holds: in the former case,
from $s\notin M_{x,y}w$ we get $s\notin Mw$, hence $s\in (Mv\setminus Mw)\setminus Mx=Mu\setminus Mx\subseteq M_{x,y}u$, so that $s \in M_{x,y}u$; in the latter case, we get $Mx\cap Mw=\emptyset$ and hence
$Mx\cap Mu=Mx\cap(Mv\setminus Mw) = Mx \cap Mv \neq\emptyset$, and therefore again $s\in M_{x,y}u$.
The inclusions $M_{x,y}u \subseteq M_{x,y}v \setminus M_{x,y}w$ and
$M_{x,y}u \supseteq M_{x,y}v \setminus M_{x,y}w$ just shown sum up to $M_{x,y}u = M_{x,y}v \setminus M_{x,y}w$; that is, $M_{x,y}\models u=v\setminus w\,$.

Consider next a literal $u \neq v \setminus w$ in $\varphi$.
If $Mu \nsubseteq Mv \setminus Mw$ then either $Mu\nsubseteq Mv$ or $Mu\subseteq Mv\wedge Mu\cap Mw\neq\emptyset$ holds; accordingly, by \lemsref{equallemma}{disjlemma}, $M_{x,y}u \nsubseteq M_{x,y}v \setminus M_{x,y}w$ holds.
On the other hand, if $Mv \setminus Mw \nsubseteq Mu$, then there exists an $s \in Mv$ such that $s \notin Mu \cup Mw$. If $s \notin Mx$, then $
s \in  M_{x,y}v$ and $s \notin  M_{x,y}u \cup M_{x,y}w$, hence $s\in M_{x,y}v\setminus M_{x,y} w$ and $M_{x,y}u\nsupseteq M_{x,y}v \setminus M_{x,y}w $; otherwise, $My\in M_{x,y}v$ follows from $s\in Mx\cap Mv$, whereas $My\notin M_{x,y}u\cup M_{x,y}w$, else either $Mx\cap Mu\neq\emptyset$ or $Mx\cap Mw\neq\emptyset$ would hold, entailing (due to the fact that $M$ models $\Xi_\varphi^\psi$) the contradiction $s\in Mx\subseteq Mu\cup Mw$; thus $My$ witnesses that $M_{x,y}u\nsupseteq M_{x,y}v \setminus M_{x,y}w$ when $s\in Mx$. Summing up, we get $M_{x,y}u \neq M_{x,y}v \setminus M_{x,y}w$ in all cases; that is, $M_{x,y}\models u\neq v\setminus w\,$.

Concerning the atom $x = \hug{y}$, it follows from $M\models\Xi_\varphi^\psi$ that
$M\models x\nsubseteq y\wedge\big(x\nsubseteq y\longrightarrow \Disj{x,y}\big)\,$, and hence $M\models x\neq\varnothing\wedge \Disj{x,y}$; therefore,
$ M_{x,y}y=My$ and $ M_{x,y}x=\hug{My}=\hug{M_{x,y}y}$ hold. That is,
$M_{x,y}\models x=\hug{y}$.

What precedes suffices to prove that $M_{x,y} \models \varphi\, \wedge\, x=\hug{y}$; since moreover $
M_{x,y}\tilde{v} \coloneqq M\tilde{v}
$
holds according to \eqref{functionMpp},
\lemsref{equallemma}{disjlemma} also yield $M_{x,y} \models \Xi_\varphi^\psi$.
\end{proof}

\begin{corollary}\label{onecorol}
Let $M$ be a model of $\varphi \wedge \Xi_\varphi^\psi$ satisfying  \eqref{notincond} for some atom $x = \hug{y}$ in $\psi$, and $\psi'$ be the conjunction of all atoms $x' = \hug{y'}$ in $\psi$ satisfying $M_{x,y}x' \neq Mx'$. Then $M_{x,y} \models \psi'$. Also, $x=\hug{y}$ belongs to $\psi'$, $M_{x,y}y=My$, and 
$M_{x,y}\models x'=x\wedge y'=y$ for each $x' = \hug{y'}$ in $\psi$.
\end{corollary}
\begin{proof}
When $x' = \hug{y'}$ in $\psi$, from $M_{x,y}x' \neq Mx'$ we get $\neg\Disj{Mx,Mx'}$. Therefore, since 
\[
M \models \bigwedge_{\substack{x'' = \hug{y''} \text{ in } \psi\\v \in \Vars{\varphi\wedge\psi}}}\big(\neg\Disj{x'',v}\longrightarrow x'' \subseteq v \big),
\]
we obtain $Mx = Mx'$; and then, by \eqref{functionM}, $M_{x,y}x' = M_{x,y}x = \hug{My}$. From $Mx = Mx'$ we also get $My = My'$, and thus $My = M_{x,y}y = M_{x,y}y'$, whence $M_{x,y}x' = \{M_{x,y}y'\}$.
The genericness of $x' = \hug{y'}$ in $\psi'$ yields $M_{x,y} \models \psi'$.
\end{proof}

\begin{lemma}\label{lem:SatXiToSatPsi}
If $\varphi \wedge \Xi_\varphi^\psi$ is satisfiable, so is $\varphi \wedge \psi\,$.
\end{lemma}
\begin{proof} Suppose that $\varphi \wedge \Xi_\varphi^\psi$ is satisfiable; then it has a $\flat$-flat model $M^{(0)}$ with $\flat>\card{\Vars{\varphi\wedge\Xi_\varphi^\psi}}\geq\card{\Vars{\varphi\wedge\psi}}$, by \lemref{lemma:rho-flat}. 
Belaboring the idea sketched at the beginning of \Sec{sec:satisfiabilityPreservation}, we consider the ordering $\prec^{M^{(0)}}$ between atoms of $\psi$ as defined in \defref{rel}
and repeatedly perform the following actions:
\begin{enumerate}[label=\roman*.]
\item\label{item:indI} choose an atom $\ell_i\defAs x_i\!\!= \!\!\hug{y_i}$ in $\psi$, minimal in regard to $\prec^{M^{(0)}}$, that does not appear in $\psi_j$ for any $j < i$;
\item define $M^{(i)} \coloneqq M^{(i-1)}_{x_i,y_i}$ and let $\psi_i$ be the collection of all atoms $x'= \hug{y'}$ in $\psi$ such that $M^{(i)}x' \neq M^{(i-1)}x'$;
\item finally, take $m$ to be the value such that $\psi_1 \wedge \cdots \wedge\psi_m = \psi$.
\end{enumerate}

We will prove by induction on $i= 1,\dots,m$ that:

\begin{enumerate}[label=${\arabic*}$\small)]
\item\label{item:indFour} Condition \eqref{notincond} holds when
$M=M^{(i-1)}$ and $y=y_i\:$. --- Thus $M^{(i)} \models \varphi \wedge \Xi_\varphi^\psi\wedge \psi_i$ holds, where $x_i=\hug{y_i}$ is a conjunct of $\psi_i$, by \lemref{oneexpress} and  \corref{onecorol}.

\item\label{item:indOne} $M^{(i)}y=M^{(k-1)}y_k$ and $M^{(i)}x=\hug{M^{(k-1)}y_k}$ hold for 
$1\leq k\leq i$ and for every atom $x=\hug{y}$ in $\psi_{k}\:$. --- Consequently $M^{(i)} \models \psi_1 \wedge \cdots \wedge \psi_i$ holds.

\item\label{item:indTwo} $\rk{M^{(i)}v}\in\{0,\dots,i\}\cup\{\flat+1,\dots,\flat+1+i\}$
holds for every $v\in\Vars{\varphi\wedge\psi}$; therefore $M^{(i)}v \notin \mathcal{V}^{\#}_{\flat}$, if we assume w.l.o.g.\ %
that $\flat > m$.

\item\label{item:indThree} The inclusion $M^{(i)}v\subseteq\big\{M^{(i)}y\wt x=\hug{y}\ \mbox{in}\ \psi_1\wedge\cdots\wedge\psi_{i}\big\}\uplus \mathcal{V}^{\#}_{\flat}$ holds for each $v\in\Vars{\varphi\wedge\psi}$.
\end{enumerate}
Thus, by \ref{item:indFour}--\ref{item:indOne}, $M^{(m)} \models \varphi\wedge \psi$ will finally hold, settling the claim of this lemma.

\medskip

{%
Notice, in passing and in view of \lemref{disjlemma}, that $\ell\prec^{M^{(i)}}\ell'\longleftrightarrow \ell\prec^{M^{(i-1)}}\prec\ell'$ will also follow from our construction and induction hypotheses for all $\ell,\ell'\in\psi$; we will hence get (omitting the model's superscript) $\ell_{k'}\!\!\nprec^M\!\!\ell_{k}$ for $1\leq k<k'\leq m$.}

\medskip

\noindent{\bf Case $i = 1$}. Concerning \ref{item:indFour}, since $M^{(0)}$ is $\flat$-flat, by \lemref{notin} condition \eqref{notincond} holds when $M=M^{(0)}$ for each atom $x=\hug{y}$ in $\psi$, and in particular when $y = y_1$. The claim \ref{item:indOne} readily follows from \corref{onecorol} in this case, as it also does, for any $i$, when $i=k\,$.

To get \ref{item:indTwo}, notice that $M^{(1)}v \subseteq M^{(0)}v \cup \{M^{(0)}y_1\}$ readily follows from the definition of $M^{(1)}$; then, since $M^{(0)}$ is $\flat$-flat, $\rk{M^{(0)}v}\in\{0,1,\flat+1,\flat+2\}$.
To get \ref{item:indThree}, rely on the inclusion  
$M^{(1)}v\subseteq \hug{M^{(0)}y_1}\cup M^{(0)}v$ just noticed, where $M^{(0)}v\subseteq \mathcal{V}^{\#}_{\flat}$ (due to $\flat$-flatness), $M^{(0)}y_1 = M^{(1)}y_1\notin\mathcal{V}_{\flat}^{\#}$ (by \ref{item:indOne} and  \ref{item:indTwo} already proved for $i=1$), and $x_1=\hug{y_1}$ belongs to $\psi_1$ (as noted under \ref{item:indFour}).

\medskip

{\noindent{\bf Case $i > 1$}.\ As regards \ref{item:indFour}, arguing by contradiction suppose that 
$M^{(i-1)}y_i\in M^{(i-1)}v$ holds for some $v\in\Vars{\varphi\wedge\psi}$. Since
$M^{(i-1)}v\subseteq\big\{M^{(i-1)}y\wt x=\hug{y}\ \mbox{in}\ \psi_1\wedge\cdots\wedge\psi_{i-1}\big\}\uplus \mathcal{V}^{\#}_{\flat}$ holds by the induction hypothesis
\ref{item:indThree} but $M^{(i-1)}y_i\in\mathcal{V}^{\#}_{\flat}$ is ruled out by  
\ref{item:indTwo}, we get $M^{(i-1)}y_i\in
\{M^{(i-1)}y\wt x=\hug{y}\ \mbox{in}\ \psi_1\wedge\cdots\wedge\psi_{i-1}\big\}$; accordingly, $M^{(i-1)}y_i = M^{(i-1)}y_j$ holds for some $j < i$, thanks to \corref{onecorol}.
Through \lemref{equallemma}, our induction ensures that $M^{(j)}\models\Xi_\varphi^\psi\,$, $M^{(j)}\models y_i = y_j$, $M^{(j)} \models x_i = x_j \leftrightarrow y_i = y_j$, and thus 
$M^{(j)}x_j = M^{(j)}x_i$ and $M^{(j)} y_j=M^{(j-1)}y_j \in M^{(j)}x_i$. Since $M^{(j-1)}y_j \notin M^{(j-1)}x_i$ holds, by hypothesis \ref{item:indFour}, the literal $x_i = \hug{y_i}$ must belong to $\psi_j$, which leads us to the sought contradiction.}

\begin{sloppypar}
As for \ref{item:indOne}, supposing the contrary, there should exist a $k$ and a least $h$ such that $k<h\leq i$ and that either $M^{(h)}y \neq M^{(h-1)}y$ or $M^{(h)}x \neq \hug{M^{(h-1)}y}$ holds for some literal $x=\hug{y}$ in $\psi_k$.  In the former case, we should have $\neg\:\Disj{M^{(h-1)}x_h,M^{(h-1)}y_k}$, which leads to the contradiction $\ell_h\prec^{M^{(0)}}\ell_k$. Therefore, we must have $M^{(h)}y=M^{(h-1)}y=M^{(k-1)}y_k$ and $M^{(h)}x \neq \hug{M^{(k-1)}y_k}=M^{(h-1)}x$. This is untenable, though: in fact, by \corref{onecorol}, 
$M^{(h)}x \neq M^{(h-1)}x$ implies $M^{(h)}x = \hug{M^{(h)}y}$, whence
$M^{(h)}x = \hug{M^{(k-1)}y_k}$ follows.
\end{sloppypar}

As regards \ref{item:indTwo}, the construction of $M^{(i)}$ gives us $M^{(i)}v \subseteq M^{(i-1)}v \cup \hug{M^{(i-1)}y_i}$ where, by the induction hypotesis \ref{item:indTwo}, either 
$0 \leq \rk{M^{(i-1)}w} < i$ or $\flat+1 \leq \rk{M^{(i-1)}w} \leq \flat +i$ holds for each  $w$
in $\Vars{\varphi\wedge\psi}$, and in particular for $w$ in $\hug{v,x_i}$. We will have
either $0 \leq \rk{M^{(i)}v} < i$ or $\flat+1 \leq \rk{M^{(i)}v} \leq \flat +i$ when
$M^{(i)}v \subseteq M^{(i-1)}v\,$; the upper bounds $i$ and $\flat+i$ must be increased by one in case
$M^{(i-1)}y_i\in M^{(i)}v\,$.
 
Finally, concerning \ref{item:indThree}, we have that:
\[\begin{array}{rcll}
M^{(i)}v &\subseteq & M^{(i-1)}v \cup \{M^{(i-1)}y_1\}\hfill\text{(by definition of $M^{(i)}$)} &\\
&\subseteq &\big\{M^{(i-1)}y\wt x=\hug{y}\ \mbox{in}\ \psi_1\wedge\cdots\wedge\psi_{i-1}\big\} \uplus \mathcal{V}^{\#}_{\flat} \cup \{M^{(i-1)}y_i\}&\\
&&\multicolumn{2}{r}{\text{(by the induction hypothesis \ref{item:indThree})}}\\
&=& \big\{M^{(i)}y\wt x=\hug{y}\ \mbox{in}\ \psi_1\wedge\cdots\wedge\psi_{i-1}\big\} \uplus \mathcal{V}^{\#}_{\flat} \cup \{M^{(i)}y_i\}&\\
&&\multicolumn{2}{r}{\text{(by the induction hypothesis \ref{item:indOne})}}\\
&=& \big\{M^{(i)}y\wt x=\hug{y}\ \mbox{in}\ \psi_1\wedge\cdots\wedge\psi_{i}\big\} \uplus \mathcal{V}^{\#}_{\flat}&\\
&&\multicolumn{2}{r}{\text{(since $M^{(i)} \models \Xi_\varphi^\psi$ and by the induction hypothesis \ref{item:indTwo}).}}\vspace{-0.65cm}
\end{array}\]
\end{proof}

\subsection{Each model of $\varphi \wedge \psi$ can be extended into a model of $\varphi \wedge \Xi_\varphi^\psi$}\label{sec:modelXtension}
Let $\varphi \wedge \psi$ be satisfiable and let $M$ be a model of it.
Consider the relation $\prec$ between set variables induced by membership
according to the rules
\begin{enumerate}[label=(\roman*)]
\item $x \prec y$ if $Mx \in My$, and
\item if $x \prec y$ and $y \prec z$ then $x \prec z$,
\end{enumerate}
so that, plainly, $\prec$ is a strict ordering. 

Extend $M$ to the set variables $\tilde{v}$ by putting, for each of them:
\[
M\tilde{v} \coloneqq \{Mx\wt x \prec v\}.
\]

To prove that $M \models \Xi_\varphi^\psi$, first notice that $Mx$ is a singleton  for all $x = \hug{y}$ in $\psi$, so that we have 
\[
M \models \bigwedge_{\substack{x = \hug{y} \text{ in } \psi\\v \in \Vars{\varphi\wedge\psi}}}\big(\neg\Disj{x,v}\longrightarrow x \subseteq v \big).
\]
Moreover $M$ models all pair of atoms $x = \hug{y}$ and $x' = \{y'\}$ in $\psi$ so that
\[
M \models \bigwedge_{x = \hug{y} \text{ in } \psi} x \nsubseteq y
\] and 
\[
M \models \bigwedge_{\substack{x = \hug{y} \text{ in } \psi\\ x' = \{y'\} \text{ in } \psi}}\big( y = y' \longleftrightarrow x = x'\big).
\]

Concerning any of the implications
\[
\neg\Disj{x,v} \longrightarrow \tilde{y} \subsetneq \tilde{v}
\]
with $x = \hug{y} \text{ in } \psi$ and $v \in \Vars{\varphi\wedge\psi}$, assume that $\neg\Disj{Mx,Mv}$; then, since $Mx = \hug{My}$, we have that $My \in Mv$ and hence $y \prec v$,
so that transitivity and strictness of $\prec$ yield $\{My'\ |\ y' \prec y\} \subsetneq \{Mv'\ |\ v' \prec v\}$; that is, $M\tilde{y} \subsetneq M\tilde{v}$.

Finally, as for the implications
\[x = y \longrightarrow \tilde{x} = \tilde{y}
\]
with $x,y \in \Vars{\varphi\wedge\psi}\,$, assume $Mx = My$. When $v \prec x$,
a tuple $v_1,\dots,v_n$ of set variables exists such that $v = v_1$, $Mv_i \in Mv_{i+1}$ holds for each $i$, and $v_n=x$; thus
$Mv=Mv_1 \in \cdots \in Mv_n = Mx = My$, and hence $v \prec y$. Analogously $v\prec x$ follows from $v\prec y$. Thus $\{v\ |\ v \prec x\} = \{v\ |\ v \prec y\}$; that is, $M\tilde{x} = M\tilde{y}$.

We have so extended a generic $M$ such that $M \models \varphi\wedge\psi$ into a model of $\varphi \wedge \Xi_\varphi^\psi$; therefore we get
\[
\models \big(\varphi(\vec y) \wedge \psi(\vec x)\big) \longrightarrow (\exists \vec z)\ \Xi_\varphi^\psi(\vec x,\vec y,\vec z\,)\,.
\]

Putting together \lemsref{quadraticCom}{lem:SatXiToSatPsi}, and the conclusion just reached, we get:
\begin{theorem}
Membership conjunctions are $\mathcal{O}(n^{2})$-expressible from \BSTbb into \BSTbbPlus. \qed
\end{theorem}

\subsection{Design and analysis of the translation algorithm}\label{sec:newTranslationAlg}

In order to prove \lemref{quadraticCom}, we provide a detailed specification of the algorithm that generates the formula $\Xi_\varphi^\psi$ out of the conjunction $\varphi \wedge \psi$.
{\small
\begin{algorithmic}[1]
\State Initialize $\Vars{\varphi \wedge \psi}$ as an empty list of set variables;
\State Initialize $\Xi_\varphi^\psi$ as an empty list of conjuncts;
\For{each set variable $x$ that appears in $\varphi$}\label{for1stBis}
	\State add $x$ to $\Vars{\varphi\wedge\psi}$;
\EndFor\label{for1endBis}
\For{each conjunct $x =\hug{ y }$ in $\psi$}\label{for2stBis}
	\State add $x$ and $y$ to $\Vars{\varphi\wedge\psi}$;
	\State add $%
	x \nsubseteq y%
	$ to $\Xi_\varphi^\psi$;
\EndFor\label{for2endBis}
\For{each conjunct $x =\hug{ y }$ in $\psi$}\label{for3stBis}
	\For{each $v \in \Vars{\varphi\wedge\psi}$}
		\State add $\big(\neg\Disj{{x},v} \longrightarrow {x} \subseteq v\big) \wedge \big(\neg\Disj{{x},v} \longrightarrow \tilde{y} \subsetneq \tilde{v}\big)$ to $\Xi_\varphi^\psi$;
	\EndFor
\EndFor\label{for3endBis}
\For{each pair $x =\hug{ y }$, $x' =\hug{ y' }$ of distinct conjuncts in $\psi$}\label{for4stBis}
		\State add $\big(y=y' \longleftrightarrow x = x' \big)$ to $\Xi_\varphi^\psi$;
\EndFor\label{for4endBis}
\For{all $x,y \in \Vars{\varphi\wedge\psi}$}\label{for5stBis}
    \State add $(x = y \longrightarrow \tilde{x} = \tilde{y})$ to $\Xi_\varphi^\psi$.
\EndFor\label{for5endBis}
\end{algorithmic}}

Adding elements to $\Vars{\varphi \wedge \psi}$ and to $\Xi_\varphi^\psi$ will require constant time if these
are implemented as lists of set variables and conjuncts.

The \textbf{for}-loop at lines~\ref{for1stBis} and \ref{for1endBis} can be performed in $\Theta(\card{\varphi})$-time, where $\card{\varphi}$ is the total length of the conjunction $\varphi$;
similarly the \textbf{for}-loop from line~\ref{for2stBis} to line~\ref{for2endBis} can be performed in $\Theta(\card{\psi})$-time.
The \textbf{for}-loop from line~\ref{for3stBis} to line~\ref{for3endBis} is iterated $\Theta\left(\card{\psi\times\Vars{\varphi\wedge\psi}}\right)$ times,
the \textbf{for}-loop at lines~\ref{for4stBis} and \ref{for4endBis} is iterated $\Theta\big(\card{\psi}^2\big)$ times, and the \textbf{for}-loop at lines~\ref{for5stBis} and \ref{for5endBis} is iterated
$\Theta\left(\card{\Vars{\varphi\wedge\psi}}^2\right)$ times.

\begin{sloppypar}
The overall time complexity then is $\Theta\big(\card{\varphi\wedge\psi} + \card{\psi\times\Vars{\varphi\wedge\psi}} + \card{\psi}^2+\card{\Vars{\varphi\wedge\psi}}^2\big)$, and since $\card{\Vars{\varphi\wedge\psi}} =  \mathcal{O}\big( \card{\varphi\wedge\psi} \big)$, we can say that $\Xi_\varphi^\psi$ can be generated in $\mathcal{O}\left(\card{\varphi\wedge\psi}^2\right)$-time.
\end{sloppypar}

\section{Related and planned work}

The nested-to-flat translation discussed so far can be seen as an instance of the quantitative approach to logical inference (cf. \cite{Hooker88,HookerF90}), as specialized to the field of Computable Set Theory (cf. \cite[Chapter~11]{COP01}).

Our translation is, in fact, meant to set up the ground
for reductions of satisfiability problems regarding sets to instances of integer programming---or, if we are to rise above the theory of hereditarily finite sets, to the language of the additive theory of cardinals (which is decidable, cf. \cite{Tar56}). 

One such reduction, where the source language embodies cardinality, is presented in \cite[Sec.~11.1]{COP01}; it can certainly be improved as we will strive to do, and we expect that it can be boosted with the treatment of rank-related constructs. Some reductions of the set-satisfiability problem to integer programming can be found in \cite{HibtiPhD}, whose line of research aimed at integrating in a single logic programming language linear programming problems and set-constraint manipulation methods, as explained in \cite{HibtiLL93}.\footnote{The endeavour of integrating cardinality constraints into constraint logic programming with sets has been carried out with a different approach, as reported in \cite{cristia_rossi_2021}.}
A technique for reducing the problem of multilevel syllogistic (cf. \cite{SCO11})
to propositional consistency testing was described in \cite{HibtiPhD} (an account of it can also be found in \cite[Sec.~11.3]{COP01}).

\medskip

In his bachelor degree thesis defended at the Univ. of Trieste
 in 2020, Mattia Furlan spotted out the
valid formulae involving Boolean difference that are shown in \Fig{ax:diffAlg}.
Let us adopt the universal closures of those formulae as the axioms of a theory based on quantificational first-order
logic with equality. These axioms characterize an algebraic variety, whose instances we provisionally dub here
\emph{difference algebras}. We have an open issue: Is every difference algebra $\mathbb{D}=(\mathcal{D},\setminus_{\cal D})$ isomorphic
to an algebra of the form $\mathbb{S}=(\mathcal{S},\setminus)$ which interprets the operator `$\setminus$' as ordinary
subtraction between sets? Here, of course, $\mathcal{S}$ must be a family of sets closed under subtraction, hence under $\cap$, because
$X\cap Y\ =\ X\setminus( X \setminus Y )$ holds for all sets $X,Y$. Perhaps, in order to settle this issue positively,
we should somehow manage to apply Stone's celebrated representation
theorem, stating that every Boolean algebra is isomorphic to a field of sets. However, we see no direct
way of relying on that theorem, because
there are difference algebras $\mathbb{D}$ whose support domain $\mathcal{D}$  fails to be closed under symmetric  difference
intended as an operation $\opair{Y}{Z}\mapsto Y\symm_{\cal D}Z$ such that, for all $X,Y,Z$ in $\mathcal{D}$,
$\begin{array}{rclcccrclcrcl}
X &=& Y \symm_{\cal D} Z &\ &\leftrightarrow&\ & X \setminus_{\cal D} (Y\setminus_{\cal D} Z) &=& Z \setminus_{\cal D} Y & \wedge & Y\setminus_{\cal D} Z &=& X\setminus_{\cal D} Z\end{array};$
moreover, it is not clear to us how one can embed a generic difference algebra into one which is a Boolean ring proper because it
enjoys this closure property.

\medskip

Today we are addressing the quest for a flat-to-nested, hopefully linear-cost, translation somehow  inverse to the one treated in this paper: that should eliminate
the equality relator from conjunctions of literals including the ones %
of multilevel syllogistic in terms of membership.

\bibliographystyle{plain} \small

\appendix
\section{A lower bound on the number of sets of a fixed integer rank}\label{app:wasLemmaOne}

Here we figure out inequalities preparatory to the proof of \Prop{prop:sizeVsharp} below.

\begin{proposition}\label{aux-sizeVsharp}
For every positive integer $n$, we have
\[
2^{2^{n}} - 2^{n} \geq 2 (2^{n} - n)
\]
or, equivalently, 
\begin{equation}\label{aux-sizeVsharpA}
2^{2^{n}} \geq 3 \cdot 2^{n} - 2 n.
\end{equation}
\end{proposition}
\begin{proof}
We prove \eqref{aux-sizeVsharpA} by induction on $n \geq 1$. For $n=1$, we have
$
2^{2^{n}} = 4 = 3 \cdot 2^{n} - 2 n.
$
For $n > 1$, by induction we have 
\[
2^{2^{n-1}} \geq 3 \cdot 2^{n-1} - 2 (n -1).
\]
Hence, 
\begin{align*}
2^{2^{n}} &= 4 \cdot 2^{2^{n-1}}\\
		  &\geq 4 \big( 3 \cdot  2^{n-1} - 2(n-1) \big) \\
		  &= 6 \cdot 2^{n} - 8n + 8 \\
		  &= \big( 3 \cdot 2^{n} - 2 n \big) + \big( 3 \cdot 2^{n} - 6 n  + 8 \big) \\
		  &> 3 \cdot 2^{n} - 2 n\,,
\end{align*}
since $2^{n} \geq 2n$ for all $n > 1$, and therefore $3 \cdot 2^{n} - 6 n  + 8 > 0$.
\end{proof}

\renewcommand{\thesection}{\arabic{section}}

\newcounter{mysection}
\setcounter{mysection}{\value{section}}
\newcounter{mytheorem}
\setcounter{mytheorem}{\value{theorem}}
\setcounter{section}{1}
\newcounter{mylemma}
\setcounter{mylemma}{\value{lemma}}
\setcounter{lemma}{0}

Next we come to a proposition which lies in the background of this paper:
\begin{proposition}\label{prop:sizeVsharp}
For every positive integer $n$, the number of well-founded sets of rank equal to $n$ is greater than or equal to $2^{n-1}$, namely 
\[
\card{\mathcal{V}^{\#}_{n}}\geq 2^{n-1}.
\]
\end{proposition}
\begin{proof}
We proceed by induction on $n \geq 1$. For $n=1$, we have $\card{\mathcal{V}^{\#}_{1}} = 1 = 2^{1-1}$. For $n > 1$, by induction we have:
\[
\card{\mathcal{V}_{n}} - \card{\mathcal{V}_{n-1}} = \card{\mathcal{V}^{\#}_{n-1}}\geq 2^{n-2}.
\]
Hence,
\begin{align*}
\card{\mathcal{V}^{\#}_{n}} &= \card{\mathcal{V}_{n+1}} - \card{\mathcal{V}_{n}} \\
	&=    2^{\card{\mathcal{V}_{n}}} - \card{\mathcal{V}_{n}} \\
	&=    2^{2^{\card{\mathcal{V}_{n-1}}}} - 2^{\card{\mathcal{V}_{n-1}}} \\
	&\geq 2 \Big( 2^{\card{\mathcal{V}_{n-1}}} - \card{\mathcal{V}_{n-1}} \Big) && \text{(by Proposition~\ref{aux-sizeVsharp}, since $\card{\mathcal{V}_{n-1}} \geq 1$)}\\
	&=    2 \big( \card{\mathcal{V}_{n}} - \card{\mathcal{V}_{n-1}} \big)\\
	&\geq 2 \cdot 2^{n-2} &&\text{(by induction hypothesis)}\\
	&= 2^{n-1}\,.  %
\end{align*} 

\vspace{-1.74\baselineskip} ~
\end{proof}

\end{document}